# Numerical simulation of high-speed penetration-perforation dynamics in layered armor shields

*M. Ayzenberg-Stepanenko[a] and G. Osharovich[b]*

[a]Ben-Gurion University of the Negev, Beer-Sheva 84105, Israel
[b]Bar-Ilan University, Ramat-Gan, 52900, Israel

**Abstract.**

Penetration models and calculating algorithms are presented, describing the dynamics and fracture of composite armor shields penetrated by high-speed small arms. A shield considered consists of hard (metal or ceramic) facing and multilayered fabric backing. A simple formula is proved for the projectile residual velocity after perforation of a thin facing. A new plastic-flow jet model is proposed for calculating penetration dynamics in the case of a thick facing of ceramic or metal-ceramic FGM materials. By bringing together the developed models into a calculating algorithm, a computer tool is designed enabling simulations of penetration processes in the above-mentioned shields and analysis of optimization problems. Some results of computer simulation are presented. It is revealed in particular that strength proof of pliable backing can be better as compared with more rigid backing. Comparison of calculations and test data shows sufficient applicability of the models and the tool.

**Introduction**

There is a set of light composite armor shields combined of "rigid" and "pliable" materials, which are widely used for contemporary protective structures vs. diverse high-speed kinetic energy projectiles (KEP), including conventional bullets. A typical structure is schematically depicted in Fig. 1. The target is a composite shield comprising a thin hard facing (F) and thick backing (B) of multilayered fabrics jointed into a matrix. The facing plate is manufactured of hard materials. Among them we note (i) steel, (ii) ceramics, and (iii) ceramic-metal composite, the so-called functionally graded material (FGM). High-strength and pliable fabric matrices (Kevlar, Dyneema, etc.) are used for the backing, fiberglass and other composites are used as well. Such shields are intended to protect light combat or cash-carrying vehicles, security doors, cabins and control rooms in boats and small ships, etc. against light arms [1].

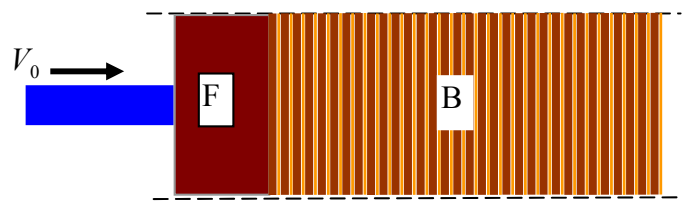

Fig.1 Impact of projectile onto composite shield

In the case of a local impact and penetration, the role of the facing is not only to decrease the impact energy, but also (and it frequently turns out to be the main factor) to subsequently spread the impact over a wide area of the backing, energy being transferred from the projectile to the protective structure. This spreading is realized due to the mushrooming of the projectile head and due to a plug of the fractured facing pushed out by the projectile onto the backing.

Hard steel shields are conventional protective structures, while ceramics have been used in the recent decades. The ability of ceramics to be used as protective and structural material against chemical, thermal and mechanical actions predetermined its wide promotion in hi-tech. Possessing a set of advantages (in comparison with metals), ceramics show weak resistance to dynamic loading, especially to local impact [2-4]. In a ceramic plate, for example, a conoid plug is formed at the free rear, and a relatively small amount of energy is absorbed in this process [5-6]. To suppress this drawback a ceramic layer is confined by metal appliqués, which prevent drastically developed fracture of free surfaces [7]. It is important to underline that under extreme conditions of high-speed penetration a brittle material can flow in an impact area as ductile, and, vice versa, a ductile material can exhibit brittle features. Therefore the same penetration models can be successfully used to describe high-speed penetration processes in metals and ceramics [8-11].

The work on ceramic-metal FGM for ballistic protection was initiated by the patent [12]. FGM aims at optimizing the performance of material components in terms of their spatial coordinates. Ceramic-metal composites are meant to combine some of the desired properties of the ceramic component, such as hardness, with those, like toughness, of the metal component. An example of the hardness profile is schematically shown in Fig. 2.

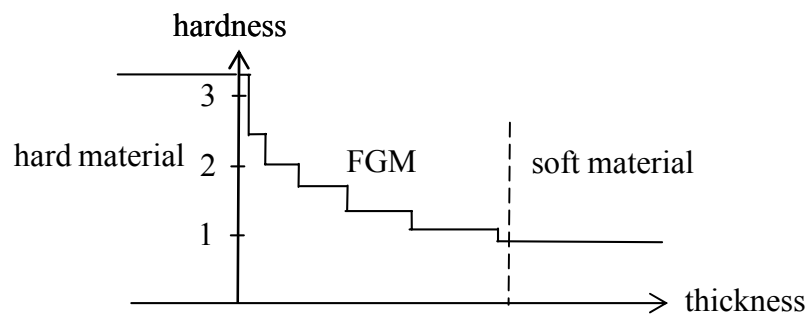

Fig. 2 Hardness profile of FGM shield

The property profiles across the FGM applique thickness should be established to maximize the resistance to penetration and to increase the stopping power of the composite target vs. high-speed projectiles. The role of the material properties in resistance to penetration in a plate varies as the projectile goes deeper and approaches the rear surface. In general outline, hardness is more important in the vicinity of the front surface, while fracture toughness, plasticity of the material and ability to resist its elongation become critical with increasing depth. For example, ship armor is usually treated by carburizing its front surface to be harder, and by specific heat treatment to increase plasticity of its rear. Some advancements concerning protective FGM design in the world-leading body, U.S. Army Laboratory, were published in [13].

There are some significant results in the mathematical modeling of high-speed penetration processes. In parallel with purely empirical dependences which have usually found applications, there are two theoretical approaches, namely:

(i) computer codes based on the general theory of continuous media and some empirical constants (see, for example, [10, 14-18]), and

(ii) semi-empirical and analytical models based on relatively simple schemes of the related processes and some experimental results at hand (see [8-11, 19 - 32]).

The most general first approach is realized in a variety of cumbersome computer codes based on general laws of mechanics, constitutive equations and some empirical constants. These codes enable a set of problems to be calculated in many important cases of impact and penetration of brittle-ductile composite structures. However, their disadvantages follow from their generality – the constitutive equations and fracture dynamics conditions have not yet been sufficiently investigated, some parameters can be obtained only by complicated experiments, which, as a rule, are too expensive.

There are many works (see, e.g., [23-28]) related to the second approach and devoted to dynamic testing and theoretical description of specific phenomena observed in fabric armors at impact and penetration regimes. However, many aspects of these complicated processes have not been adequately studied and no finally completed theory has been developed to describe dynamic fracture in metal/ceramic-fabric structures under impact. All this makes it difficult to use theoretical methods in the topical problem of optimizing composite protection. It could be also seen that there is no considerably developed theory of penetration into FGM.

In the present paper we develop and link models (empirical, analytical and numerical) that describe several penetration stages in order to examine the stopping power of composite targets with three types of facing – thin hard steel, thick ceramics or FGM, and with multiply fabric backing. We use the likely assumption that the backing does not influence the facing; the developed models are related to successive independent stages: perforation of the primary armor and penetration into the secondary one. Then we bring together the developed algorithms and elaborate a common computer tool. Calibration of the tool has been done on the basis of a comparison of calculation and test results implemented by the Rocket Systems Division (RSD) at the Israeli Military Industries Co.

First, the designed formula for perforation of a thin backing is presented.

## 1 DYNAMIC PUNCTURE OF A THIN METAL FACING

Mathematical modeling of such a process at high-speed impact results in empirical formulas (see, e.g., [22, 25, 29]) including parameters of credible rheology of the material. In this paper a more simple formula based on energy consideration and on a single empirical is designed as a result of data processing. We use the tests conducted by the RSD that consisted of seven series of a number of shots in each by M-16 and AK-47 bullets onto steel plates of various thicknesses and hardnesses. The scheme of tests with M-16 shots is shown in Fig. 3 (*a*). For this bullet we have: $m = 3.6$g, $d = 5.62$ mm (caliber), $L_0 = 15$mm (the length of an effective cylinder that is to be used below). Impact and residual velocities, $V_0$ and $V_r$, are measured.

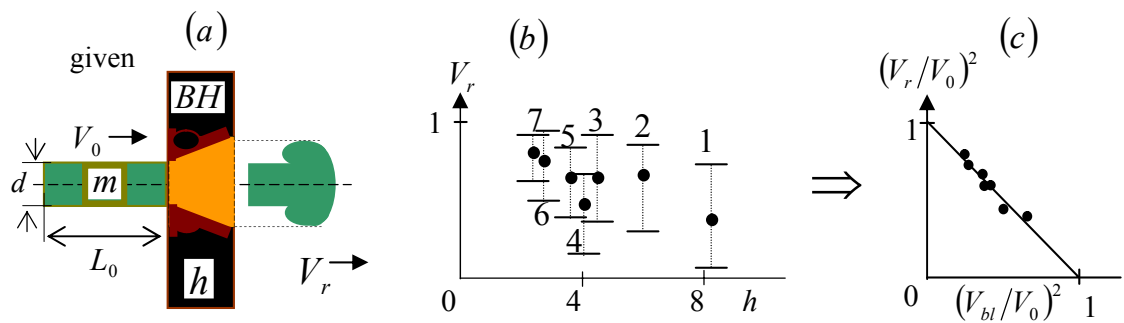

Fig. 3 Scheme of the test $(a)$, and data processing: $(b) \Rightarrow (c)$.

In Table 1 results are presented of 7 series of shots by M-16 bullets. Target parameters, BH (Brinell hardness number), and thickness, $h$, were invariable for the current series. There were 5 ÷10 shots in each series with the impact velocity within the range of $V_0 \sim 955 \div 1025$ m/s, while the ranges for plate parameters are BH ~ 480 ÷700 and $h \sim 2.4 \div 8.9$ mm. Output of test results with series numbers is shown in Fig. 3 (*b*).

Table 1

| series number | $h$ (m) | BH (KPa/mm²) | $V_0$ (m/s) | $V_r$ (m/s) measured | $V_{rc}$ (m/s) calculated | $V_r / V_{rc}$ |
|---|---|---|---|---|---|---|
| 1 | 0.0082 | 505 | 1012 | 391 | 383 | 0.98 |
| 2 | 0.0060 | 505 | 998 | 639 | 596 | 0.94 |
| 3 | 0.0048 | 525 | 970 | 624 | 655 | 1.05 |
| 4 | 0.0045 | 700 | 967 | 471 | 503 | 1.06 |
| 5 | 0.0041 | 595 | 971 | 628 | 652 | 1.04 |
| 6 | 0.0032 | 595 | 969 | 742 | 732 | 0.99 |
| 7 | 0.0029 | 595 | 973 | 802 | 762 | 0.95 |

As a result of the energy approach applied to the interaction between the projectile and the thin metal facing, the model yields finite formulas for the energy release at the perforation (and as a consequence, the ballistic limit velocity, $V_{bl}$) and for the residual velocity after perforation. The latter then plays the role of the initial impact velocity for the backing.

In the situation that key thermo-physical parameters of the explored fast dynamic process are unknown we tend to design data processing as simply as possible.

Let us consider the ratio of bullet kinetic energy $E_{kin} = mV_0^2/2$ and work of plastic strains $W_p = \gamma \sigma_Y \int_Q \varepsilon(q) dq$, where $\gamma$ is a part of kinetic energy absorbed by plastic resistance, $\sigma_Y$ is the yield limit and $\varepsilon(q)$ is the strain distribution within the deformed domain $Q$.

Then we reconstruct $W_p$ using data at hand: $W_p = f \cdot HB \cdot h \cdot d^2$ where $f$ is the fitting factor. Our aim is to evaluate factor $f$. Equating $E_{kin}$ and $W_p$ we obtain

$$V_{rc} = \sqrt{V_0^2 - V_{bl}^2}, \quad V_{bl}^2 = f \cdot BH \cdot h \cdot d^2 / m, \qquad (1.1)$$

where $V_{bl}$ is the ballistic limit.

Formula (1.1) provides the best approximation to data if factor $f = 2.47 \cdot 10^7$: a good coincidence can be seen between the results presented in the two last columns in Table 1 and Fig. 1 (c) (in the latter test data marked by circles). It was also obtained: $f = 2.95 \cdot 10^7$ for an AK-47 bullet (its initial parameters are: $m = 0.0097$ kg, $d = 0.0076$ m). The residual diameter of bullets after perforation was evaluated as the diameter of the outcome at the plate backing: $d_r = 1.27d$ (M-16) and $d_r = 1.33d$ (AK-47). The data determined at this stage are used below as the input required for calculation of penetration processes in the considered composite shields. As to the exit mass after perforation, $m_r$, the tests discussed above result in a significant dispersion ($m_r \sim 0.6 \div 1.6$). The reason is the difficulty in accounting for a huge amount of exited debris.

## 2   PLASTIC FLOW MODEL OF PENETRATION INTO A THICK FACING

Below we present calculations based on a jet plastic-flow model enabling all residual parameters of high-speed penetration to be obtained with the same accuracy.

A hydrodynamic quasi-steady-state model of high-speed penetration by metal KEP into thick targets is developed, as a constrained flow of the projectile and target materials with regard to plastic flow resistance. The formulation is related to permanently improved jet models of high-speed penetration by a KEP, with a half-century history. In the conventional classification (see, e.g., [19, 22]) the process of high-speed penetration by a KEP into a thick target is subdivided into several stages. Among them, the quasi-steady-state stage of the projectile motion within the target dominates. A major part of the kinetic energy of the projectile is consumed at this stage, which determines the main penetration parameters: penetration depth, projectile erosion and crater size.

This first (and simplest) hydrodynamic model [30] – a collision of two jets of ideal fluids – is asymptotically exact because hydrodynamic factors become dominant with increase in the penetration velocity, $V$. In this sense, the ratio $\rho V^2/\sigma_Y$ is decisive ($\rho$ and $\sigma_Y$ are the density and yielding limit of the target material). However, for regular ballistic velocities, one to two thousand meters per second, it is not high enough to permit neglecting the strength factor. The resistance of materials to penetration was introduced into the jet model by Alekseevski [31] and Tate [32]: two strength factors, $\sigma_p$ and $\sigma_t$, related to resistance of projectile and target materials $P$ were added into the "modified" Bernoulli equation for jet collision. This version is still in use (see [33-34]) for estimation of crater depth in targets of plastic and brittle materials and for evaluation in tests of the above-mentioned strength factors. At the same time, its essential drawback is that the model provides no way to determine the crater geometry and the projectile shape. The latter is of essential practical importance most notably for composite armor. In [8] and [11] the jet model was successively improved. Firstly, in [8], the movement is taken into account of backward jets in the direction opposite to penetration (see Fig. 3) realized under the condition of detached flows. Such an improvement enables the mentioned parameters to be evaluated. Secondly, two new strength factors, $\sigma_{p+}$ and $\sigma_{t+}$, are introduced into the modified Bernoulli equations for backward jets. Whereas $\sigma_p$ and $\sigma_t$ are confirmed to satisfy the experimental data concerning the crater depth, the newly introduced parameters are theoretically defined from the expression for the plastic work. The latter is obtained in [11] based on the scheme of proportional strain of backward jet materials, which allows this work to be defined in terms of the initial and final parameters of the flow. Lastly, a thermo-viscoplastic penetration problem was considered in [35], in which the shear localization phenomenon and melt wave motion were described. The resistance to shear in the molten layer decays almost to zero, it results in separation of the projectile-target materials motion with negligible shear stresses in the interface and validates the modified Bernoulli equation for plastic jets.

In this paper, the model [11] is adapted to the thick facing of the FGM. First of all we explain the geometrical scheme of the model presented in Fig.3. Section A–A is the free surface, O is the stagnation point (the point of jets collision), section B–B is the current stationary penetration state, in which penetration zone I (area of target and projectile interaction) bounded by radius $r = R$, $R - R_1$ and $R_1 - R_0$ are thicknesses of target and projectile backward jets respectively; in zone II ($r > R$) the target is immobile; $V$ is the current velocity of the uneroded part of the projectile, $V_{t-}$ is the penetration velocity, $V_{t+}$ и $V_{p+}$ are velocities of backward jets. Sub-indices "+" and "–" are taken for jets flowing in the penetration direction and in the opposite to it, excluding $V_{t\rightarrow}$, which is directed to the free surface. It is related to the conventional description of the jet model, in which the axial coordinate moving with the stagnation point is used – the target jet runs against the projectile one.

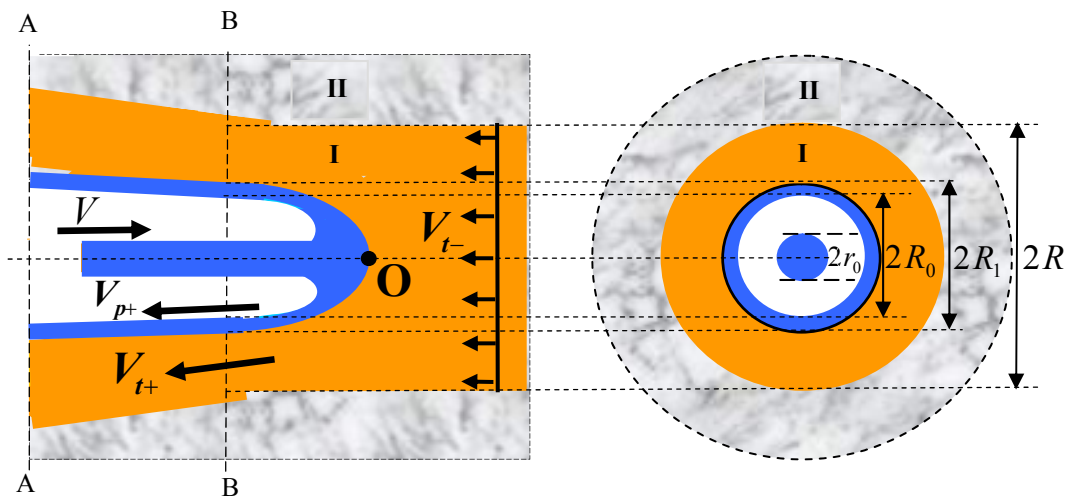

Fig.4 Scheme of the flow of detached jets

Then we present the mathematical description of the problem related to the above-mentioned scheme. First, the steady-state process is considered. In this way, all the main parameters of the geometry are obtained: the boundary between immobile and fluid target material, $R$, the radius of the projectile mushroom cup, $R_1$, the crater radius, $R_0$, as well as the erosion rate of the projectile. Then, the deceleration of the projectile, penetration depth, $P$, and crater volume, $Q_c$, as functions of time are determined based on the step-by-step numerical algorithm.

First, the steady-state coordinate system relation is

$$V_{p-} = V(t) - V_{t-}, \qquad (2.1)$$

where $V_{p-}$ is the projectile erosion velocity, $V(t)$ is the current projectile velocity, and $V_{t-}$ is the penetration velocity. Then, the Bernoulli equation for the pressure at the stagnation point O is

$$\rho_t V_{t-}^2 + 2\sigma_t = \rho_p V_{p-}^2 + 2\sigma_p, \qquad (2.2)$$

where $\sigma_t$ and $\sigma_p$ are experimental constants for the target and projectile. Note that equations (2.1) and (2.2) are present in the Alekseevski-Tate model which is true thanks to the common stagnation point. As confirmed by a set of tests (see, e.g., [22]) $\sigma_t \sim 3 \div 6 \sigma_Y^{(t)}$, $\sigma_p \sim 1.5 \div 3 \sigma_Y^{(p)}$, where $\sigma_Y^{(t)}$ and $\sigma_Y^{(p)}$ are static yielding limits of target and projectile materials, respectively.

Let in the case of the FGM facing, the dynamic strength factor of target depends on the penetration depth $P$: $\sigma_t = \sigma_t(P)$, while $\sigma_t$ = const remains for a homogeneous target material.

In addition to the usual formulation, we introduce the Bernoulli equation for each backward jet too. In doing so, we modify the equation to take into account the specific work of plastic strain, $\sigma_t^*(P)$ and $\sigma_p^* \sim$ const, for each jet during the flow. That is, two additional equations are introduced:

$$\rho_t V_{t-}^2 + 2\sigma_t(P) = \rho_t V_{t+}^2 + 2\sigma_{t+}(P), \qquad (2.3)$$

$$\rho_p V_{p-}^2 + 2\sigma_p = \rho_p V_{p+}^2 + 2\sigma_{p+}, \qquad (2.4)$$

which serve below to obtain the geometrical parameters of the target and projectile flows. The additional strength factors, $\sigma_{t+}(P)$ and $\sigma_{p+}$ are to be obtained by the analysis of the plastic resistance in backward jets.

For estimation of these specific works, we now base our considerations on the scheme of the proportional strain of the materials, the Mises plasticity condition and the associated law of the plastic strain. Possible hardening and influence of temperature on the "global" flow of the jets are neglected. In the considered case, as a result of the deformation, a cavity of radius $R_0$ arises in the target, and the projectile head transforms to the back jet, whose internal radius is the cavity of radius $R_0$. The final strain is defined by these values and the axial elongations which are assumed to be independent of the radial coordinate. These conditions lead to the following expressions for the specific works of plastic strain averaged over the radial coordinate:

$$(\sigma_{t+}, \sigma_{p+}) = \sigma_Y^{(t,p)} \frac{1}{\lambda\sqrt{3}} \int_0^\lambda \sqrt{\ln^2(1+1/x) + 3\ln^2 \lambda_z}\, dx$$

$$(\sigma_{t+}): \lambda = \frac{R^2}{\lambda_z R_1^2}, \lambda_z = \frac{R^2}{R^2 - R_1^2}; \quad (\sigma_{p+}): \lambda = \frac{R^2}{\lambda_z R_1^2}, \lambda_z = \frac{r_0^2}{R_1^2 - R_0^2} \qquad (2.5)$$

where $\sigma_Y^{(t)} = \sigma_Y^{(t)}(P)$.

The additional (to those of energy) equations of the model are the momentum equation:

$$R^2(\sigma_t + \rho_t V_{t-}^2) = (R^2 - R_1^2)\rho_t V_{t+}^2 + r_0^2(\sigma_p + \rho_p V_{p-}^2) + (R_1^2 - R_0^2)\rho_p V_{p+}^2 \qquad (2.6)$$

and the incompressibility equations:

$$R^2 V_{t-} = (R^2 - R_1^2) V_{t+}, \qquad (2.7)$$

$$r_0^2 V_{p-} = (R_1^2 - R_0^2) V_{p+}. \qquad (2.8)$$

The system (2.1) – (2.8) serve for the determination of the seven unknowns $V_{t-}, V_{t+}, V_{p-}, V_{p+}, R, R_1, R_0$, where $R_0(t)$ and $R_1(t)$ are the crater and projectile mushroom cup radii, respectively.

Note that in the momentum equation (2.6) shear stresses are neglected on the boundary of the flow, $r = R$, and on the interface between the projectile and target materials (the surface containing point O). It is based on results in the study of the localization phenomenon [35].

If the steady-state solution is obtained, then the transient problem is to be solved. The current (residual) length $L_r(t)$, uneroded mass $m(t)$ of the projectile and its velocity are

$$L_r(t) = L_0 - \int_0^t V_{p-} dt, \quad m(t) = \pi r_0^2 \rho_p L_r(t); \quad \dot{V}(t) = -\sigma_p/m(t), \quad V(0) = V_0, \quad (2.9)$$

where $L_0$ and $V_0$ are initial parameters. The current crater depth, $P(t)$, and the current crater volume $Q(t)$ (as the volume of the cavity of radius $R_0$) are

$$P(t) = \int_0^t V_{t-} dt, \quad Q(t) = \pi \int_0^t R_0^2(t) \frac{dP(t)}{dt} dt \quad (2.10)$$

The system (2.1) – (2.11) completely determines the considered problem. These formulas are enough to calculate the penetration velocity and the crater depth, while $\sigma_{t+}(P)$ and $\sigma_{p+}$ (2.5), serving for the crater diameter and volume determination, are a priori unknown and are obtained within the calculation process on the basis of the iteration method.

A calculation algorithm and a computer program have been built on the basis of a step-by-step finite difference algorithm. The time step $\Delta t$ is chosen from the required accuracy condition. At the current time $t$ we have all the needed parameters calculated before and set

(i) The "initial" data for the iteration process is $\sigma_{t+}(P) = \sigma_Y^{(t)}(P)$, $\sigma_{p+} = \sigma_Y^{(p)}$.

(ii) From (2.1) and (2.2) we obtain

$$V_{t-} = V(t)\left(1 - \sqrt{1-\gamma(1-\beta)}\right)\gamma^{-1}, \quad V_{p-} = V(t)\left(\sqrt{1-\gamma(1-\beta)} - \alpha\right)\gamma^{-1} \quad (\alpha \neq 1).$$
$$V_{t-} = 0.5V(t)(1-\beta), \quad V_{p-} = 0.5V(t)(1+\beta) \quad (\alpha = 1); \quad (2.11)$$
$$\alpha = \rho_p/\rho_t, \quad \gamma = 1-\alpha, \quad \beta = 2(\sigma_t - \sigma_p)/\rho_p.$$

(iii) Using (2.3) and (2.4) we find the speeds of backward jets:

$$V_{t+} = \sqrt{V_{t-}^2 + 2(\sigma_t - \sigma_{t+})/\rho_t}, \quad V_{p+} = \sqrt{V_{p-}^2 + 2(\sigma_p - \sigma_{p+})/\rho_{pt}}.$$

(iv) All radii are obtain from (2.6)–(2.8), while (2.5) yields new $\sigma_{t+}(P)$ and $\sigma_{p+}$.

At the following iteration calculations repeat taking into account these new values. The iteration process is stopped if the required convergence is reached, i. e. a preassigned error is proved in obtaining $\sigma_{t+}(P)$ and $\sigma_{p+}$ in the process of the coupled calculation of the geometry and the plastic resistance. After that the calculation of (2.9) and (2.10) results in the new penetration parameters and transition occurs to the next time layer, $t + \Delta t$. Calculations are stopped if one of conditions below is satisfied:

$$\beta \geq 1, \quad V \leq 0, \quad V_{t-} \leq 0, \quad V_{p-} \leq 0, \quad L_r \leq 0, \quad \text{Im}\{V_{t-}, V_{p-}, V_{t+}, V_{p+}\} \neq 0$$

The described algorithm was realized in a computer program which works, as calculations show, very fast. Below we describe some calculations conducted for "experimental" projectile-target pairs, for which a huge amount of test data is presented in [36, 37]. These pairs consist of three types of plates manufactured of armor steel HzB20 (below: $t_1$), Steel 52 ($t_2$) and Steel 37 ($t_3$) and of two types of cylindrical projectiles manufactured of steel C110W2 (below, $p_1$) and high-density metal Densimet D17 ($p_2$). In Table 1 the data related to material properties is presented, while in Table 3 the ratios range of calculated values of final crater depth, $P$, of final crater radius, $R$, and of final crater volume, $Q$, to experimental those are presented for some $p/t$ pairs and in the impact velocities range of $1000 \div 3000$ m/s. Good quantitative coincidence of results can be seen. In Fig.5 and 6 normalized penetration parameters vs. time are depicted: $\overline{P} = P/L_0$, $\overline{L_r} = L_r/L_0$, $\overline{V_{(\ )}} = V_{(\ )}/V_0$, $\overline{Q} = Q/q$, where $q$ is the initial volume of the projectile. The shown data and similar results obtained for other pairs enable comparison between the simple jet model and the one developed here: at relatively small velocities $V_0$ the projectile erosion speed significantly exceeds the penetration speed (~ twice at $V_0$ = 1000 m/s). If $V_0$ increases these speeds are drawn together approaching the hydrodynamic limit equal to $\frac{1}{2}V_0$.

Table 2. Parameters of projectiles and targets [36, 37]

| projectile: | d (mm) | L/d | $\rho$ (kg/m$^3$) | $\sigma_Y^{(p)}$ (MPa) | $\sigma_p$ (MPa) |
|---|---|---|---|---|---|
| $p_1$ – Steel C110W2 | 2.5, 4.3, 5.4 | 10 | 7850 | 770 | 1100 |
| $p_2$ – Densimet D17 | 2.8, 6.0 | 10.4 | 17000 | 750 | 1550 |
| target: | | | $\rho$ (kg/m$^3$) | $\sigma_Y^{(t)}$ (MPa) | $\sigma_t$ (MPa) |
| $t_1$ – Steel HzB20 | – | – | 7850 | 1000 | 5175 |
| $t_2$ – Steel St-52 | – | – | 7850 | 610 | 4400 |
| $t_3$ – Steel St-37 | – | – | 7850 | 500 | 3450 |

Table 3: Theoretical-to-experimental data ratios

| | $p_1/t_1$ | | | $p_1/t_2$ | | | $p_1/t_3$ | | | $p_2/t_3$ | | |
|---|---|---|---|---|---|---|---|---|---|---|---|---|
| $V_0$ | P | D | Q | P | D | Q | P | D | Q | P | D | Q |
| 1000 | – | – | – | 0.29 | 0.91 | 0.44 | 0.81 | 0.93 | 0.80 | 1.06 | 1.02 | 0.96 |
| 1200 | 0.86 | 0.88 | 0.92 | 0.82 | 0.92 | 0.73 | 0.97 | 0.96 | 0.95 | 1.17 | 1.15 | 1.05 |
| 1500 | 1.03 | 0.90 | 1.06 | 0.95 | 0.96 | 0.87 | 1.10 | 0.98 | 1.00 | 1.06 | 1.14 | 1.08 |
| 1800 | 1.07 | 0.91 | 1.03 | 0.92 | 1.00 | 0.91 | 1.00 | 0.98 | 1.00 | 0.95 | 1.12 | 1.18 |
| 2100 | 1.00 | 0.91 | 1.12 | 0.88 | 1.00 | 0.90 | 0.93 | 1.01 | 1.03 | 0.91 | 1.09 | 1.16 |
| 2400 | 0.94 | 0.91 | 1.21 | 0.87 | 1.02 | 0.90 | 0.91 | 1.03 | 1.08 | 0.87 | 1.03 | 1.03 |
| 3000 | – | – | – | 0.87 | 1.03 | 0.95 | 0.87 | 1.06 | 1.15 | 0.83 | 0.94 | 0.89 |

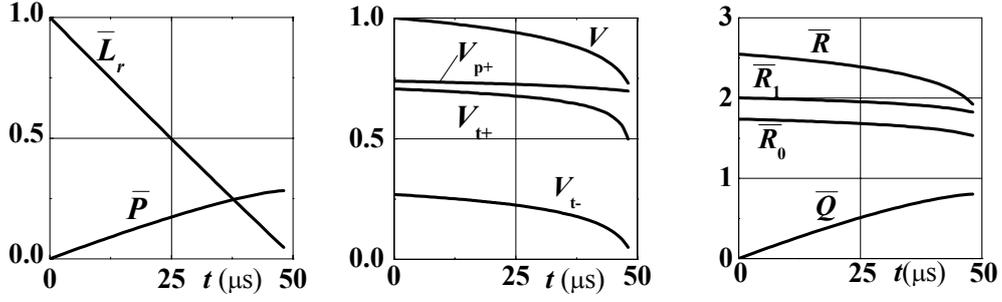

Fig. 5 Penetration parameters vs. time: $p_1/t_1$ – pair ($r_0 = 2.7$mm), $V_0 =1500$ m/s

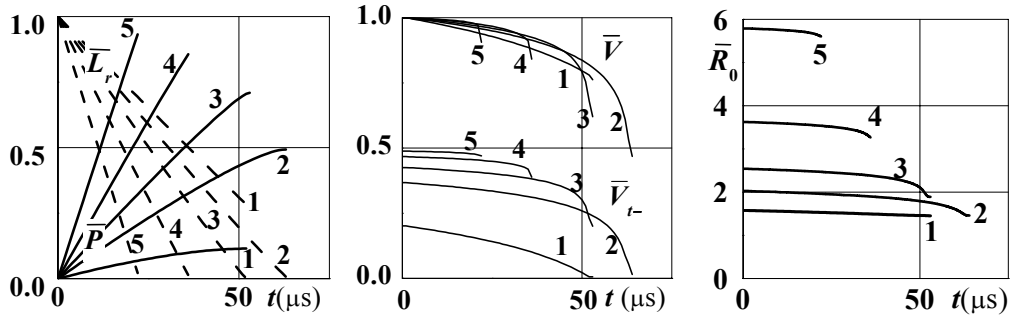

Fig. 6 Penetration parameters vs. time: $p_1/t_3$ – pair ($r_0 = 2.7$mm), curves 1, 2, 3, 4, 5 – $V_0 = 1000, 1500, 2000, 3000, 5000$ м/с.

## 3  PENETRATION MODEL FOR MULTIPLY FABRIC BACKING

The 3D axisymmetric unsteady-state fracture dynamics problem is examined for a multiplied fabric normally impacted by a rigid projectile. The fabric consists of thin plies, firm in tension, connected by well-deformed adhesive, whose main role is to distribute stresses over plies. We consider the main assumptions of the designed model:

- the plies themselves are deformed as thin membranes; the bending stresses in plies are too small with respect to tension stresses and are not taken into account;
- the adhesive is an inertionless solid in tension-compression and shear modes,
- a ply fails if the tensile strain achieves a given limiting value;
- normal/shear flaking in the adhesive occurs if stresses reach the given limiting values;

All the used stress-strain diagrams $\sigma = \sigma(\varepsilon)$ in a ply and in the adhesive are arbitrary. The above-listed assumptions enable the related mathematical model to be built. In contrast to conventional models, in which backing is a single plate with so-called "effective" geometrical and physical parameters, we build a governing system of equations on the basis of "discrete" formulation: ply motion and its interaction with the neighbors are described for each ply. Here we omit all technical details of the discretization procedure including the manner of mesh reconstruction along with progressive projectile motion, peculiarities of penetration/perforation mechanics (plies fracture and delamination cracks, plug formation, perforation event, etc.). Some points related to the mathematical model and the computer tool can be found in [15, 28, 38-40].

The designed calculation tool consists of two main blocks (below – I and II):

I  Perforation of the facing.
I.1 Thin metal facing: $V_r$ and $d_r$ are established with the use of experimental data at hand and fitting formula (1.1).
I.2 Thick metal, ceramic or FGM facing: a subroutine with algorithm (2.11) runs outputting $V_r$, $d_r$ and $m_r$.
II Penetration of composite shields: The input data induced consisting of material constants, projectile-impact data obtained in block I, and initial 3D mesh steps. The main subroutine runs for calculation of system (3.1) – (3.3). Within each time-step the mesh is rebuilt depending on the current fracture and delamination dynamics. The calculations are stopped if the projectile residual velocity is equal to zero, or the complete perforation occurs of the given target.

The tool, as was examined, has fast CP-time. Calibration of the tool has been completed by comparing the calculation results with the tests conducted by the RSD. In this way, some effective parameters were evaluated that are presented in the model, but not sufficiently confirmed by purposeful tests (generally speaking, the firmly stable data exist only for static loading).

## 4  CALCULATION RESULTS AND ANALYSIS

The results of some simulations of composite shield penetration are presented below. First, metal-fabric composites are under our examination: subroutines I.1+II run in the tool. Linear $\sigma - \varepsilon$ relations were used for backing, Young moduli $E_t$ (tangential in-ply), $E_s$ and $E_z$ (tangential and normal directions in the adhesive) and corresponding limiting stresses: $\sigma_{t,\lim}$, $\sigma_{s,\lim}$ and $\sigma_{z,\lim}$, are inputed.

A set of displacement configurations ("photos") of a shield vs. time is depicted in Fig. 7. Here and below the ply position corresponds to boundaries between light and dark strips. The input parameters are: projectile – AK-47 bullet, $V_0 = 740$ m/s; facing – steel, $h = 0.004$m, BH = 500 (projectile residual parameters after facing perforation are: $V_r = 567$ m/s $m_r = 0.003$ kg, $d_r = 0.0072$ m); backing – Kevlar-29: $m_c = 19.4$ kg/m$^2$, $E_t = 70$ GPa, $\sigma_{t,\lim} = 1.4$ GPa, $E_s = 0.5$ GPa, $\sigma_{s,\lim} = 25$ MPa, $E_z = 2.0$ GPa, $\sigma_{z,\lim} = 0.1$ GPa. The number of plies in the package is: $N_p = 40$. The output penetration parameters: time, $t$, depth, $P$, and projectile velocity, $V$, are presented below each "photo".

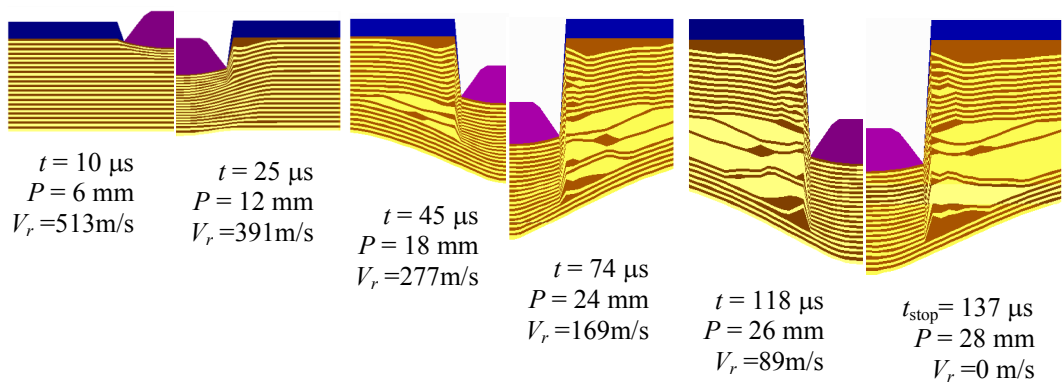

Fig. 7 Configurations of composite target vs. time

It can be seen that delaminations (beginning at time between $t = 25$ μs and $45$ μs) separate the backing into three parts: the upper package (plies n ~ 1,…,21) punctured relatively fast with small strains occurring in the vicinity of the penetration area; the second essentially delaminated part (n ~ 22,…,32) characterized by propagation of disc cracks in the adhesive with time and resulting in significant spreading of strain energy into the periphery of a very pliable intact third part (n ~ 33,…,40), which stopped the bullet at the final penetration stage.

With the optimization problem in mind, some related calculations have been performed. First, shields of the steel facing were examined. In Fig. 8 two "photos" of penetrated shields are shown. The previous facing and backings are examined (excluding $E_t$ for backings, which varied as shown in the table inside the figure). The impact parameters are: M-16 bullet, $V_0 = 1000$ m/s ($V_r = 567$ m/s $m_r = 0.011$ kg, $d_r = 0.0095$ m).

| n | $E_t$ GPa | P mm | $N_p^*$ | $W_c$ % | $V_r$ m/s |
|---|---|---|---|---|---|
| 1 | 50 | 18.4 | 20 | 100.0 | 0 |
| 2 | 70 | 21.9 | 26 | 100.0 | 0 |
| 3 | 100 | 17.7 | 40 | 95.0 | 167 |
| 4 | 140 | 17.4 | 40 | 93.5 | 193 |

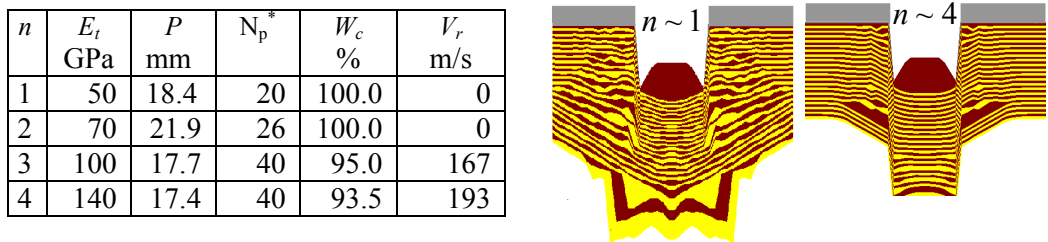

Fig. 8 Comparison of penetration in shields of different Young moduli of plies

The four last columns of the table consist of: the final penetration depth $P$ (in the case of perforation, $P$ is fixed at time, $t_{perf}$, when all plies are fully broken out and the drag force is equal to zero), volume of perforated plies, $N_p^*$, amount of energy consumed at the penetration event, $W_c$, and residual velocity, $V_r$ (nonzero at the perforation event). The photos on the right related to the first and the last rows in the table show the projectile stopped ($n \sim 1$, $t_{stop} = 116$ μs) and the shield perforated ($n \sim 4$, $t_{perf} = 69$ μs). These results confirm the known fact: more rigid armors can possess less stopping power. Note that the related concept of the so-called super-plastic protective structures (SPPS) can be found in [40].

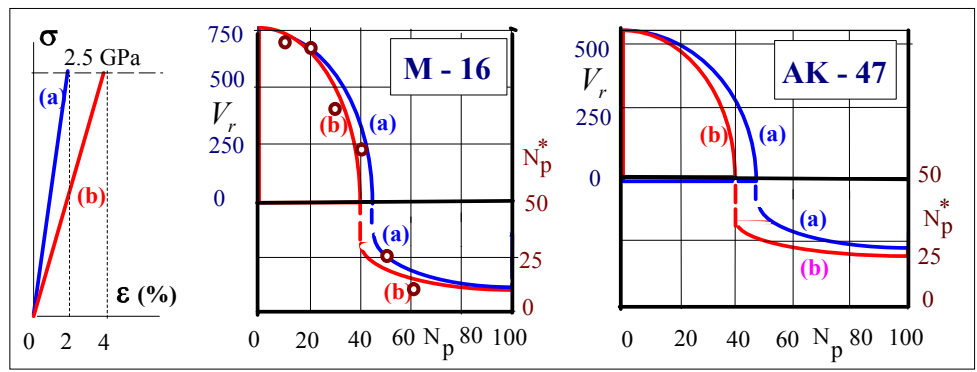

Fig. 9 Stopping power of Kevlar-29 and Kevlar-49 backings

In Fig. 9 the stopping power is compared of shields with "rigid" – Kevlar-49 (*a*) and "pliable" – Kevlar-29 (*b*) backings vs. M-16 and AK-47 bullets. Stress-strain diagrams of these materials are shown on the left. It can be seen that the stopping power of the pliable fabrics proves to be better than that of the rigid ones (a). Simulations of penetration processes were conducted with backings of varied ply number: $N_p$ = 0,...,100, residual velocities after facing perforation correspond to those shown at $N_p$ = 0. It can be seen that 40 plies of Kevlar-49 are enough to stop both types of bullets, while 45 plies (M-16) and 48 plies (AK-47) of Kevlar-49 are required for this. On the right vertical axis, $N^*_p$, the volume of penetrated plies ("trauma") is pointed in the case when the bullet is stopped by the target. As can be seen, an AK-47 bullet makes the deeper trauma. Circles seen in the figure for an M-16 bullet correspond to test data obtained by RSD for Kevlar-29 backing manufactured of 10, 20, 30, 40, 50 and 60 plies. Comparison of tests and calculations shows the good prediction ability of the tool.

In the last example, ceramic and FGM facings are examined (subroutines I.2+II run in the tool) in the case of penetration into semi-infinite targets. The strength parameters of the ceramic facing needed in the jet model are constants: $\sigma_Y^{(t)}$ = 500 MPa, $\sigma_t$ = 1.5 GPa, while those of the FGM linearly decrease till 250 MPa and 750 MPa, respectively, over the length equal to 10mm. After that they remain constants. The density is constant: $\rho_t$ = 3000 kg/m³. An AK-47 bullet plays the role of the projectile. The aim of simulations was to obtain output data needed to run the subroutine for calculation of the backing penetration process. In Fig. 10 (a) related to perforation of the facing, asterisks are the values of output parameters at time $t = t_{10}$ (when the penetration depth, $P$ = 10 mm, is achieved in the ceramic facing), while dark circles are similar indicators in the FGM case. The time of the penetration stop is marked by $t_{stop}$. The output velocity in the ceramic case is greater (by ~ 17%) than in the FGM case. At the same time the output diameter is greater (by ~ 15%) in the latter case (the residual masses turn out to be practically the same). In Fig. 10 (b) these parameters are used as input ones in calculations of penetration into the Kevlar-29 backing with $N_p$ = 30. It is obtained in both cases: $N^*_p$ = 15, although the penetration depth turns out slightly greater in the case of the ceramic facing.

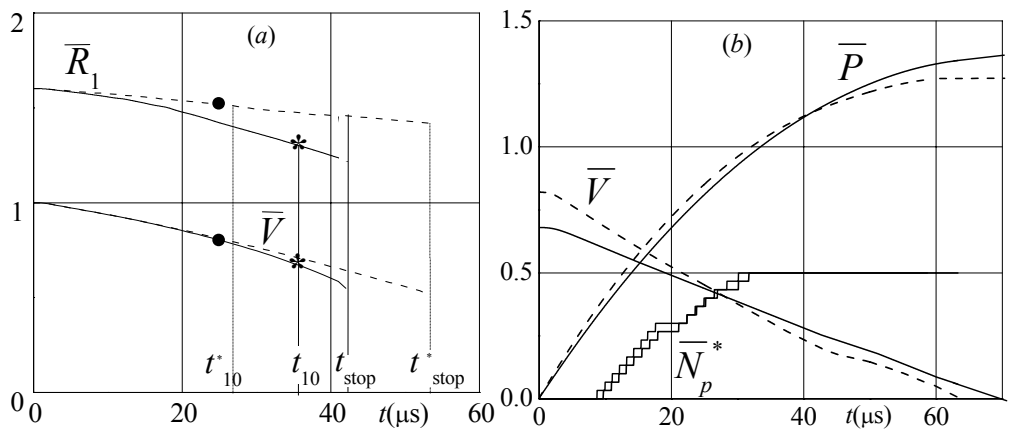

Fig. 10 Penetration into composite shields: solid curves – ceramic facing, dashed curves – FGM facing. (*a*) perforation of the facing, (*b*) penetration into the backing.

Note that the presented above examples related to the optimization problem have tentative character and aimed to show facilities of the model and calculating tool.

ACKNOWLEDGEMENTS: The authors are indebted to Dr. N. Farber for the technical collaboration and the permission to use the test data, and to Prof. N. Frage for his valuable consulting related to FGM materials.**References**